\documentclass[letterpaper,12pt]{article}
\usepackage{tabularx} 
\usepackage{amsmath}  
\usepackage{graphicx} 
\usepackage[margin=1in,letterpaper]{geometry} 
\usepackage{cite} 
\usepackage[final]{hyperref} 
\hypersetup{
	colorlinks=true,       
	linkcolor=blue,        
	citecolor=blue,        
	filecolor=magenta,     
	urlcolor=blue
}
\usepackage{amssymb}
\usepackage[T1]{fontenc}
\usepackage[utf8]{inputenc}
\usepackage{mathptmx} 
\usepackage{authblk}
\usepackage{tabto}
\usepackage{subcaption}
\usepackage{float}
\usepackage{chngcntr}

\begin{document}

\title{Assessing Guaranteed Minimum Income Benefits and Rationality of Exercising Reset Options in Variable Annuities}

\author[1]{Riley Jones} 
\author[2]{Adriana Ocejo \thanks{CONTACT A. Ocejo. Email: amonge2@uncc.edu}}
\affil[1,2]{Department of Mathematics and Statistics, University of North Carolina at Charlotte, 9201 University City Blvd., Charlotte, NC}

\date{\vspace{.5cm}\today}
\maketitle

\begin{abstract}
A variable annuity is an equity-linked financial product
typically offered by insurance companies. The
policyholder makes an upfront payment to the insurance company and, in return, the insurer is required to make a series of payments starting at an agreed upon date. For a higher
premium, many insurance companies offer additional guarantees or options which protect
policyholders from various market risks. This research is centered around two of these options: the guaranteed minimum income benefit (GMIB) and the reset
option. The sensitivity of various parameters on the value of the GMIB is explored, particularly the guaranteed payment rate set by the insurer. Additionally, a critical value for future interest rates is calculated to determine the rationality of exercising the reset option. This will be able to provide insight to both
the policyholder and policy writer on how their future projections on the performance of the stock
market and interest rates should guide their respective actions of exercising and pricing variable
annuity options. This can help provide details into the value of adding options to a variable
annuity for companies that are looking to make variable annuity policies more attractive in a
competitive market.
\vskip 1.5in
\end{abstract}
{\bf Keywords:} guaranteed minimum income benefit; Monte Carlo; pricing; reset option; variable annuity.
\pagebreak

\section{Introduction}

\tab A variable annuity is a long-term, tax-deferred product, whose funds are equity-linked from the time of the initial payment until the annuitization date (the accumulation period). The initial payment is invested into sub-accounts made up of mutual funds and other investments. The growth of the investments during the accumulation phase affects the payout of the annuity at the annuitization date (often at retirement). This product is designed to provide post-retirement income.

While this product is targeted at providing financial security throughout retirement, there is a large amount of risk inherent. This risk stems largely from the performance of the markets from which the value of the annuity is derived. If the markets perform poorly over the accumulation period, an individual could have a post-retirement income significantly less than expected. With retirement being something that few people are willing to risk, it is important to be able to offer something that reduces the risk of the variable annuity. The most common way to protect the annuity balance from poor investment performance is the inclusion of a guaranteed minimum benefit when the contract is underwritten. For a good introduction to different types of investment guarantees we refer the reader to Hardy (2003).

In fact, when insurance companies began to include guaranteed minimum benefits in their variable annuity products in the late 1990’s, there was a large growth in the number of polices sold (Drexler, Plestis, \& Rosen, 2017). This made variable annuities a more attractive option because it reduced the level of risk in these policies to policyholders. Today, guaranteed minimum benefit options are very common with variable annuities. According to Drexler et al. (2017) in 2016, 76\% of policyholders chose to purchase a guarantee when the option was available with their variable annuity. With a decline in the number of pension plans and other traditional forms of retirement plans, many people are looking into less traditional ways to be financially secure through retirement. Since variable annuities are a long-term investment which can have a very low risk (with a guaranteed minimum benefit), they are a great option for retirement.

Guaranteed minimum benefits can come in many forms; however, the four main types are Guaranteed Minimum Withdrawal Benefits (GMWB), Guaranteed Minimum Death Benefits (GMDB), Guaranteed Minimum Accumulation Benefits (GMAB), and Guaranteed Minimum Income Benefits (GMIB). Some works related to the pricing of these specific guarantees in isolation are (Milesvsky \& Salisbury, 2006), (Milevsky \& Posner, 2001), (Shevchenko \& Luo, 2016), and (Marshall, Hardy, \& Saunders, 2010), respectively. With a few exceptions, closed-form formulas are usually not available and numerical methods, such as Monte Carlo, have to be used to price these guarantees. When the computational time is a concern, particularly when dealing with large portfolios of variable annuities, some more advanced and eficient methods have been proposed, see for instance (Gan, 2013), (Gan \& Valdez, 2018), (Doyle \& Groendyke, 2019).

While (Bauer, Kling, \& Russ, 2008) and (Bacinello, Millossovich, Olivieri, \& Pitacco, 2011) amongst others have created a pricing framework for guaranteed minimum benefits in general, there is little research into GMIB. 
Furthermore, while companies have advanced pricing tools and methods to price the cost of these policies, current research indicates that these policies are typically underpriced (Marshall et al., 2010).
Marshall et al. (2010) study the value of the GMIB as a function of the fee rate $c$ charged for the rider, whereas our paper is centered around the value of the GMIB as a function of the guaranteed annual payment rate $g$, which has a large impact on the probability of exercising the benefit base and on the value of the guarantee. A GMIB could have many additional options added to it; in this paper, it essentially provides the policyholder with two options at the annuitization date: annuitize the accumulated value of investments at prevailing rates or annuitize a guaranteed amount at a set rate $g$ (determined at the onset of the contract). This research will specifically 
focus on the effect of factors such as the guaranteed payment rate $g$, the fee structure, and the volatility parameter on the value of the GMIB.

The reset option, as defined within this paper, provides a third option which allows the policyholder to defer the annuitization date to a later time. This is useful if the policyholder is not in need of a payment at the annuitization date and thinks the policy will gain value over the next year. We do not price the reset option, instead we are interested in the rationality of exercising the reset option upon the annuitization date based on future interest rate expectations.

The rest of the paper is organized as follows. In Section \ref{sec:framework}, the pricing framework of a GMIB is presented to determine fair values of the guaranteed annual payment rate $g$ given different levels of a fee rate $c$. In Section \ref{sec:results} we use Monte Carlo methods to analyze the probability of exercising the GMIB for different levels of the fee and guaranteed rates, to find the fair value of the guarantee rate, as well as how these quantities are affected by the fee structure and the volatility. Additionally, the reset option is analyzed to find critical values for future interest rates which will determine the rationality of exercising the reset option. It is important to note that the reset option is not considered in the pricing of the GMIB. Concluding remarks appear in Section \ref{sec:conclusion}.

\section{Pricing framework} \label{sec:framework}

\tab We consider a single premium variable annuity with a GMIB. This rider guarantees the policyholder the maximum of the the benefit base and the investment account at the annuitization date $T$.
The payoff of the GMIB is given by:
\begin{equation}
P(T) = max[BB(T), S_f(T)]
\end{equation}
where $BB(T)$ represents the value of the benefit base and $S_f(T)$ represents the investment account with all fees deducted. Throughout this paper, a value of $T = 20$ will be used, indicating a 20-year accumulation period, and fees will be deducted annually.  

The initial premium, $S(0)$, is invested in a fund account with market value $S(t)$ defined on a complete probability space $(\Omega, \mathcal F, \mathbb P)$. We assume that under the risk-neutral probability measure $\mathbb{Q}$, the investment account $S(t)$ \textit{before} fees are deducted follows
\begin{equation*}
    dS(t)=rS(t)dt+\sigma S(t)dW(t),
\end{equation*}
where $r>0$ is the continuously compounded risk-free interest rate, $\sigma>0$ is the market volatility, and $W=(W(t))_{t\geq 0}$ is a standard Brownian motion.
At the annuitization date, the value of the benefit base can be expressed as:
\begin{equation}
    BB(T) = S(0)(1 + r_g)^T \, g \, a_{20}(T)
\end{equation}
where $r_g>0$ is the guaranteed annual rate, $a_{20}(T)$ is the market value of a twenty-year annuity with payments of \$1 beginning at time $T$, and $g$ is the guaranteed annual payment rate specified at the beginning of the contract.

Note that if $g$ is priced fairly, then $g$ should be the multiplicative inverse of $a_{20}(T)$; however, this relationship is affected by the prevailing interest rate at time $T$. Since the interest rate at time $T$ is not known (in practice) at any time before $T$, its value has to be approximated. It is often the case that $g$ is set so conservatively that $g\,a_{20}(T) < 1$.

While some companies make annuity payments monthly, payments for the annuities priced in this paper will be paid annually for 20 years. There is flexibility for the term of the policy; however, 20 years is selected because current publicly available Social Security data show that the average life span after retirement at age 65 is 19.3 years for males and 21.7 years for females in the United States (Social Security Administration).

For the investment account, insurance companies typically deduct an annual fee that is a percentage of the benefit base. The amount deducted from the investment account every year can be expressed by:
\begin{equation} \label{eq:f_1}
    f_1(n) = cS(0)(1 + r_g)^n,\  n = 1, 2, \ldots, T.
\end{equation}
With this, the fee that will be deducted every year is known at the onset of the contract. This fee structure is similar to that in (Marshall et al., 2010). Another method for calculating the amount of the annual fee can be given by:
\begin{equation} \label{eq:f_2}
    f_2(n) = f_1(n)\, g\, a_{20}(T),\  n = 1, 2, \ldots, T
\end{equation}
With (\ref{eq:f_2}) for an annual fee, two extra terms are included: $g$ and $a_{20}(T)$. While these terms should multiply to equal one, resulting in $f_2(n) = f_1(n)$, this is often not the case. The value of $a_{20}(T)$ will change as the future expectation of the interest rate at time $T$ changes. The use of these two fee amounts and the effects they have on the pricing of the GMIB is discussed in Section \ref{sec:results}. However, $f_1(n)$ is industry standard.
By the risk-neutral valuation approach, the value if the GMIB can be expressed by:
\begin{equation}
    V(g) = \mathbb{E}_{\mathbb Q}[(1 + r)^{-T}P(T)].
\end{equation}

In insurance, an \textit{equivalence principle} is used to determine fair rates (Olivieri \& Pitacco, 2015). In our setting, we define the fair guaranteed annuity payment rate as the value of $g = g^*$ such that:
\begin{equation}
    V(g^*) = S(0),
\end{equation}
that is, the risk-neutral value of the GMIB equals that of the total investment (in this case $S(0)$).
If $V(g) > S(0)$, then the insurance company is undercharging for the GMIB. Likewise, if $V(g) < S(0)$, then the insurance company is overcharging for the GMIB.

For the reset option, the same modeling process is used. The only difference is that the model is extended by 1-year to include the projections from time $T$ to $T+1$. This option can be expressed by setting the new terminal time to $T' = T + 1$. While the interest rate is at the same level from time 0 to $T$, its value is changed for the year following the initial annuitization date, but still at a constant level. This allows for results on how the change in the interest rate will affect the values of the benefit base and the investment account. The interest rate which results in a benefit base value equal to the investment account at time $T'$ will be referred to as the critical interest rate value $r^*$ from $T$ to $T'$. If $S_f(T) > BB(T)$, then the reset option should be exercised if the policyholder expects future interest rates to be below $r^*$ (and vice versa).  


\section{Numerical results and discussion} \label{sec:results}

\tab In this section, the value of the GMIB is given as a function of the guaranteed annual payment rate $g$ to determine the fair rate $g^*$ for varying fee rates $c$ and critical values for the reset option are given. Additionally, discussion is provided into the sensitivity to other parameters of the model such as the volatility of the market, and the fee structure implemented. Unless otherwise stated, the fee structure follows that of $f_1(n)$ and the following parameters are used in the model: $S(0) = \$100,000$, $T = 20$, $\sigma = 10\%$, $r = 5\%$ and $r_g = 5\%$.

\begin{figure}[hbt!]
  \centering
  \includegraphics[width=0.65 \linewidth]{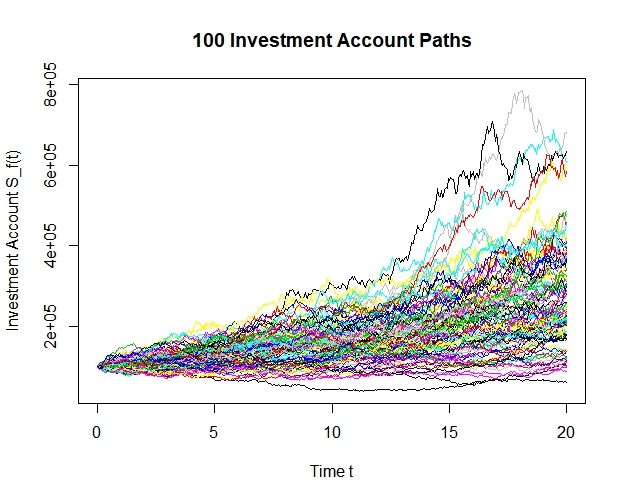}
  \caption{100 Investment account paths.}
  \label{fig:paths}
\end{figure}

A Monte Carlo simulation approach is used in the pricing of the GMIB to undertake the analysis. 
While Bauer et al. (2008) and Bacinello et al. (2011) also use Monte Carlo methods in a more general framework, we are interested in the effect of parameters specific to GMIB in the value of such guarantee. Figure \ref{fig:paths} shows 100 realizations of the investment account path given $S(0) = 100,000$, $\sigma = 10\%$ and $r = 5\%$. The most robust model run in this paper considers 200,000 simulations to estimate the values of $V(g)$ for each small increment of $g$. \\*

\begin{figure}[H]
  \centering
  \includegraphics[width=.65\linewidth]{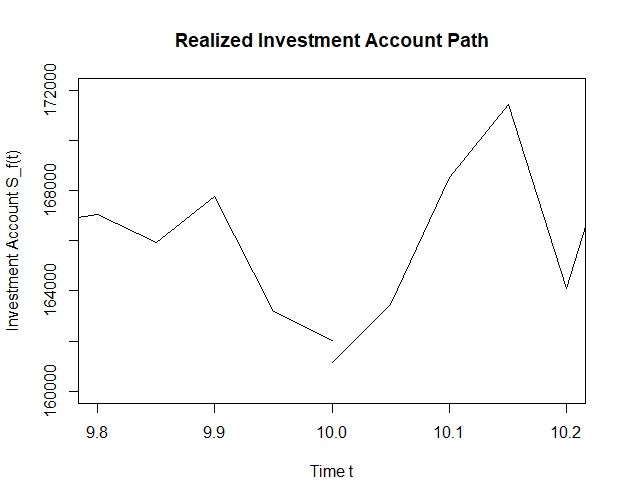}
  \caption{Deduction of annual fee.}
  \label{fig:gap}
\end{figure}

\pagebreak
It is important to note that the paths are not continuous. 
Each path is stopped at the end of each year and the fee is deducted. Then the motion continues but starting at the new account value. Figure \ref{fig:gap} is a close-up chart of one realization of the investment account. The gap represents the fee amount that was deducted from the investment account, in this case $f_1(10)$.

\subsection{\textit{Probability of exercising the GMIB option}} \label{subsec:prob}

\begin{figure}[H]
  \centering
  \includegraphics[width=.75\linewidth]{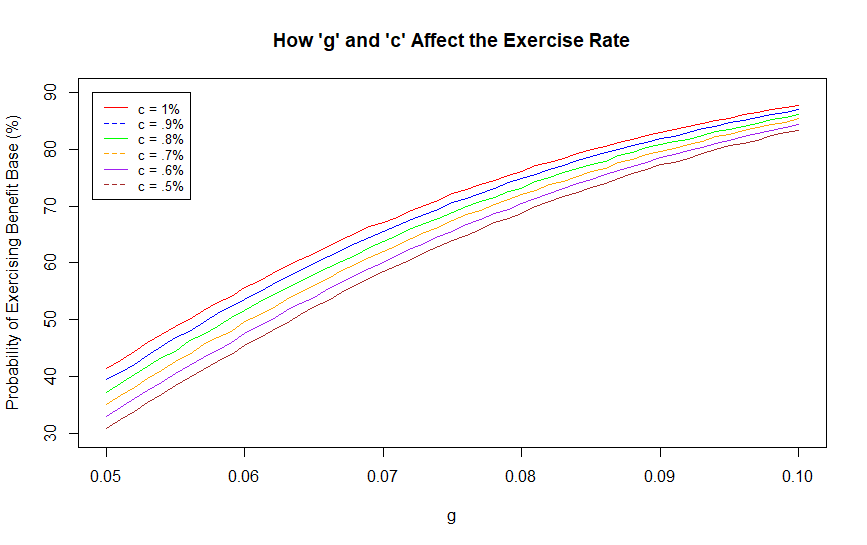}
\caption{Probability of exercising the benefit base.}
\label{fig:prob_ex}
\end{figure}

An important point of consideration when offering a guaranteed minimum benefit is understanding the probability that the benefit will be exercised. 
Our simulations show that the effect on the likelihood of payoff of small changes of $g$ is larger than that of $c$ for the same percentage increase.
Figure \ref{fig:prob_ex} shows the probability that the GMIB will be exercised for different levels of $c$ and $g$. The value of $g$ is given between 5\% and 10\% because these values correspond with a realistic range of future interest rates from 0\% to 9\% (c.f. Table 1 in (Marshall et al., 2010)). If interest rates were to exceed 9\%, then a value of $g$ higher than 10\% should be explored. Also, the fee rate ranges from .5\% to 1\% because this is current industry standard.  
Figure \ref{fig:prob_ex} is generated from points calculated for given levels of $g$ and $c$. It is connected by lines at each level of $c$ to show a trend. For each level of $c$ there are 51 values of $g$ generated on equidistant intervals from .05 to .10. 

From this graph we can see that both $g$ and $c$ have a positive relationship with the probability of exercising the benefit base. As both $c$ and $g$ increase, the probability that the benefit base is exercised increases. However, within the range of current industry standard, we can see that $g$ has a much larger impact on the probability of exercising the benefit base than $c$ does. A change from $g = .05$ to $g = .1$ results in a change of approximately 50 percentage points while the change from $c = .5\%$ to $c = 1\%$ results in a change of approximately less than 10 percentage points.

\pagebreak
Now that the effect $g$ and $c$ have on the probability of exercising the benefit base is given, we can shift focus to determining what the fair value of $g$ will be.

\subsection{\textit{$V(g)$ and fair values of $g$} \label{subsec:fair_g}}

\begin{figure}[h!]
  \centering
  \includegraphics[width=.65\linewidth]{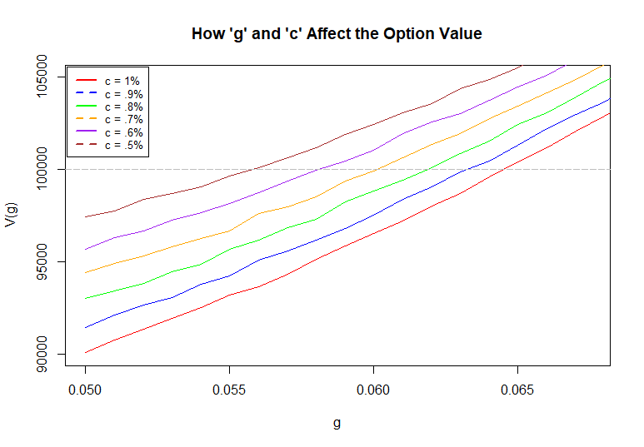}
\caption{V(g) to determine fair $g$ values.}
\label{fig:fair_g}
\end{figure}

It is important for both the insurance company and the policyholder to know what the fair value of $g$ is when signing the policy. Using the same data that was computed in the previous section, the fair value of $g$ is now calculated for the same six fee rates varying from .5\% to 1\% on equidistant intervals. Recall that the fair value of $g$ is the $g^*$ such that $V(g^*) = S(0)$. 
The $V(g)$ values are presented in Figure \ref{fig:fair_g} for each level of $g$ and $c$. 
This is done for the same 306 combinations of $g$ and $c$ calculated in the previous section. The values are then connected by lines by the fee value to show the trend of the data.
As expected, 
as the $g$ value increases, so does $V(g)$, which makes the GMIB more valuable. However, as the fee rate $c$ charged increases, $V(g)$ decreases making the value of the GMIB less valuable. 

\pagebreak
The most critical point of consideration in this section is the initial value of the investment account (100,000). When $V(g)$ is equal to 100,000, then the GMIB option is fairly priced. This is seen by the intersection of each line with the grey dotted line. If $V(g)$ is greater than 100,000, then the insurance company is undercharging for the option. On the other hand, if $V(g)$ is less than 100,000, then the insurance company is overcharging for the GMIB.

It is important to note that none of the combinations of $g$ and $c$ calculated resulted in a $V(g)$ value of exactly 100,000. However, the strict monotonicity of $V(\cdot)$ apparent in the chart
imply that there exists some value of $g$ for each level of $c$ such that $V(g)$ equals to 100,000. 
Table \ref{tbl:bins} presents an estimate of $g^*$ for various levels of $c$.
These fair values of $g$ are best estimates given the parameters of the model. If the parameters of the model were to change, then these values would change as well. Additionally, there is a degree of uncertainty with each fair value given. The degree of uncertainty varies with the number of iterations run in the model. If only a few number of iterations were run, then each fair value of $g$ would have a very large variance level associated with it. Since 200,000 iterations were run at each point, the fair $g$ values presented in Table \ref{tbl:bins} have a relatively smaller variance level. However, future work should be done into calculating the exact certainty of these values.

\begin{table}[H]
\begin{center}
\caption{Fair values of $g$ given $c$}
\label{tbl:bins} 
\begin{tabular}{|c|cccccc|}
\hline
$c$ & 0.005 & 0.006 & 0.007& 0.008& 0.009 & 0.01 \\
\hline
$g^*$ & 0.0558 & 0.0581 & 0.0601 & 0.0619 & 0.0633 & 0.0645 \\
\hline
\end{tabular}
\end{center}
\end{table}

\subsection{\textit{Fees given by $f_1(n)$ vs. $f_2(n)$}}

As stated at the beginning of this section, the fee structure used in sections \ref{subsec:prob} and \ref{subsec:fair_g} follows that of $f_1(n)$. Now, the fee structures of $f_1(n)$ and $f_2(n)$ are compared to find the probability of exercising the benefit base and the fair level of $g$. 
The relationship between $f_1(n)$ and $f_2(n)$ can be seen as $f_2(n)$ = $f_1(n) * g * a_{20}(T)$. Since the interest rate is assumed to be constant at 5\% from 0 to $T$, the value of $a_{20}(T)$ will be constant as well. So the effect $a_{20}(T)$ has on the total fee deducted is the same for any level of $g$ and $c$. In a model where interest rates vary, $a_{20}(T)$ would vary as well. This would make its effect on the fee structure more complex.

\begin{figure}[H]
\begin{subfigure}{.5\textwidth}
  \centering
  \includegraphics[width=1\linewidth]{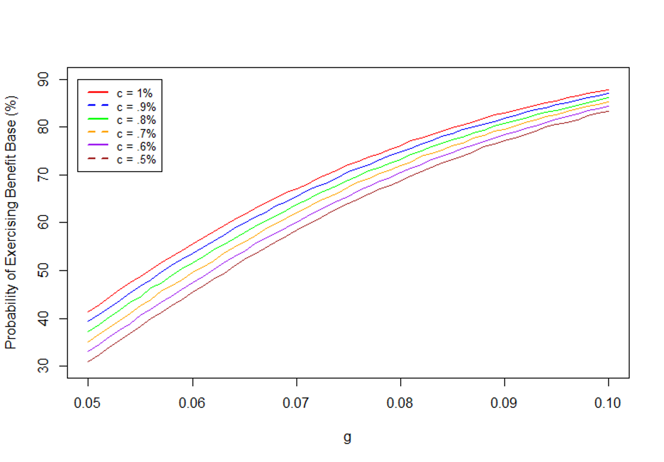}
  \caption{$f_1(n)$}
\end{subfigure}
\begin{subfigure}{.5\textwidth}
  \centering
  \includegraphics[width=1\linewidth]{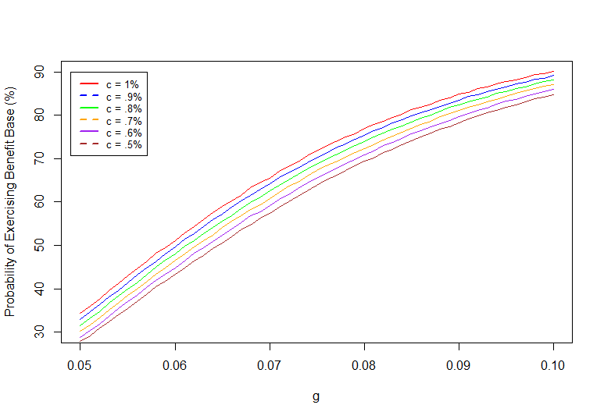}
  \caption{$f_2(n)$}
\end{subfigure}
\caption{Probability of exercising the benefit base given the fee structure.}
\label{fig:prob_givenfee}
\end{figure}

Figure \ref{fig:prob_givenfee} shows the probability of exercising the benefit base for each fee structure. While the results follow similar structures, we can see that $f_1(n)$ has more variance in $c$ at a lower level of $g$ and converges to a level of less variance in $c$ at a higher level of $g$. Meanwhile, the variance in $c$ for $f_2(n)$ appears to be constant for all $g$. Additionally, $f_1(n)$ has less change in the probability of exercising the benefit base as $g$ increases. For example, for $c = 1\%$, the probability of exercising the benefit base changes from 41\% to 88\% (47 percentage points) as $g$ goes from .05 to .1 under $f_1(n)$. Under $f_2(n)$, the probability of exercising the benefit base changes from 34\% to 90\% (56 percentage points) as $g$ goes from .05 to .1.

\begin{figure}[H]
\begin{subfigure}{.5\textwidth}
  \centering
  \includegraphics[width=1\linewidth]{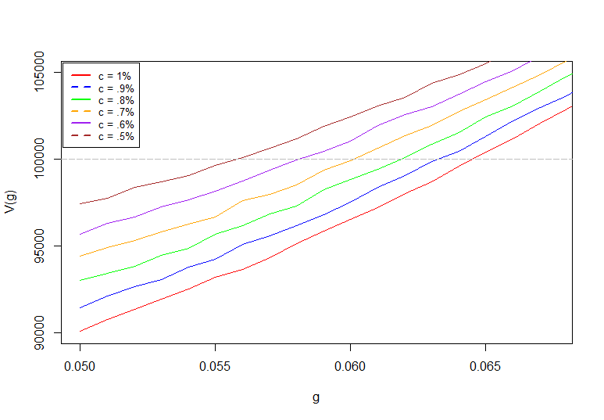}
  \caption{$f_1(n)$}
\end{subfigure}
\begin{subfigure}{.5\textwidth}
  \centering
  \includegraphics[width=1\linewidth]{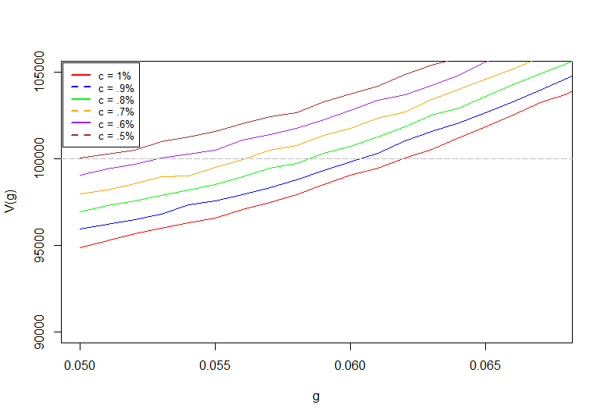}
  \caption{$f_2(n)$}
\end{subfigure}
\caption{V(g) values given the fee structure.}
\label{fig:value_givenfee}
\end{figure}

The fee structure also has a significant impact on the value of $V(g)$. As can be seen in Figure \ref{fig:value_givenfee}, there is an upward shift in the $V(g)$ values from $f_1(n)$ to $f_2(n)$. Additionally, the impact that $c$ has on $V(g)$ is less in $f_2(n)$ than $f_1(n)$ because the values are much closer together under $f_2(n)$. This results in fair values of $g$ under $f_2(n)$ which are much lower than the fair values calculated under $f_1(n)$. This makes sense because $g$ is a factor in $f_2(n)$ but not in $f_1(n)$. If the additional $g$ term is lower, then the amount being deducted from the investment account annually will be less. This will make the option more valuable, which translates into a higher $V(g)$ value. This is seen by the upward shift between the graphs.

\subsection{\textit{The effect of volatility}}

Another assumption held to be true in the previous sections is that the volatility is a constant at 10\%. So, how does changing this assumption affect the overall results? In order to test this, the model was run at a volatility level of 2\%, 10\% and 20\%.

\begin{figure}[H]
\begin{subfigure}{.5\textwidth}
  \centering
  \includegraphics[width=1\linewidth]{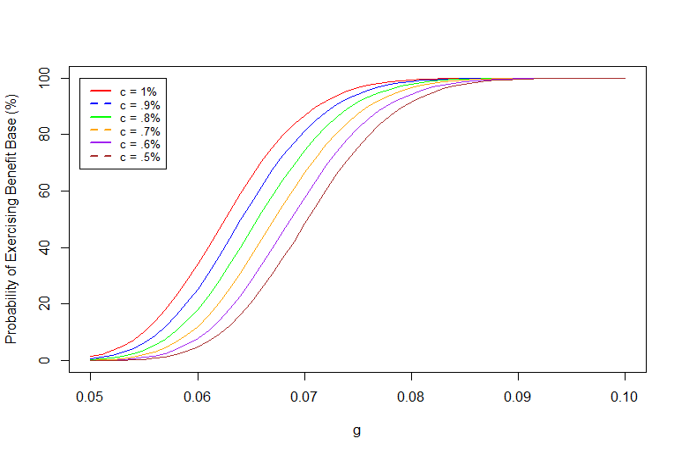}
  \caption{$\sigma$ = 2\%}
\end{subfigure}
\begin{subfigure}{.5\textwidth}
  \centering
  \includegraphics[width=1\linewidth]{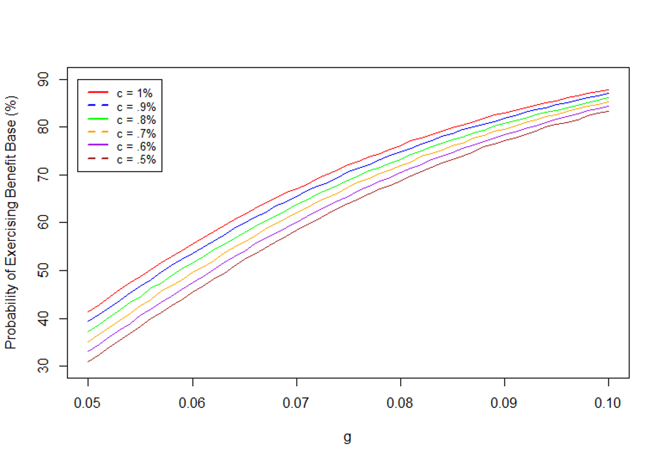}
  \caption{$\sigma$ = 10\%}
\end{subfigure}
\begin{subfigure}{\textwidth}
  \centering
  \includegraphics[width=.5\linewidth]{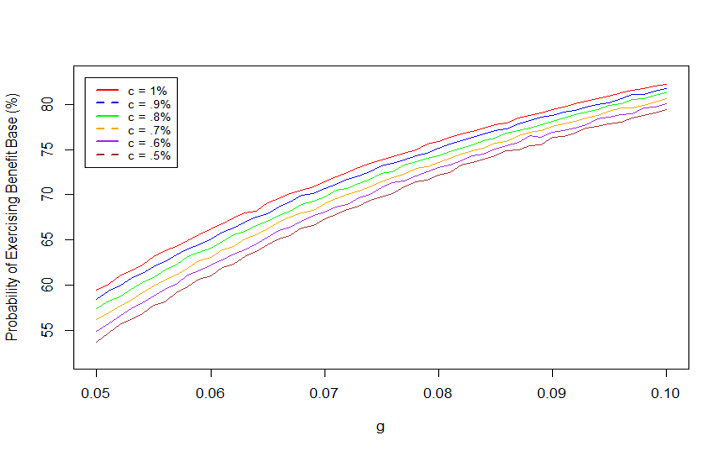}
  \caption{$\sigma$ = 20\%}
\end{subfigure}
\caption{Probability of exercising the benefit base for different volatility levels.}
\label{fig:prob_givenvol}
\end{figure}

The probability of exercising the benefit base at each of these volatility levels is given in Figure \ref{fig:prob_givenvol}. In this, we can see that the volatility has a very significant impact on the probability of exercising the benefit base. With a low level of volatility, the data appears to fit a logarithmic distribution and varies from 0\% to 100\%. As the volatility increases, the probability of exercising the benefit base is contained in a smaller interval and the data follows more of a quadratic (almost linear) function. Additionally, the lines are less smooth with increased volatility which indicates more variance in the individual results.

\begin{figure}[H]
\begin{subfigure}{.5\textwidth}
  \centering
  \includegraphics[width=1\linewidth]{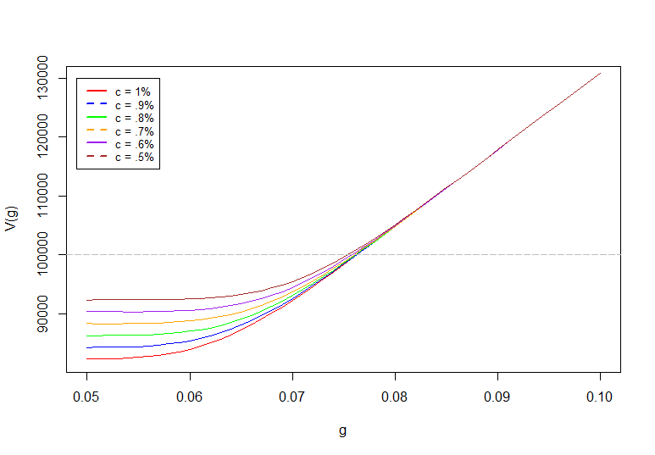}
  \caption{$\sigma$ = 2\%}
\end{subfigure}
\begin{subfigure}{.5\textwidth}
  \centering
  \includegraphics[width=1\linewidth]{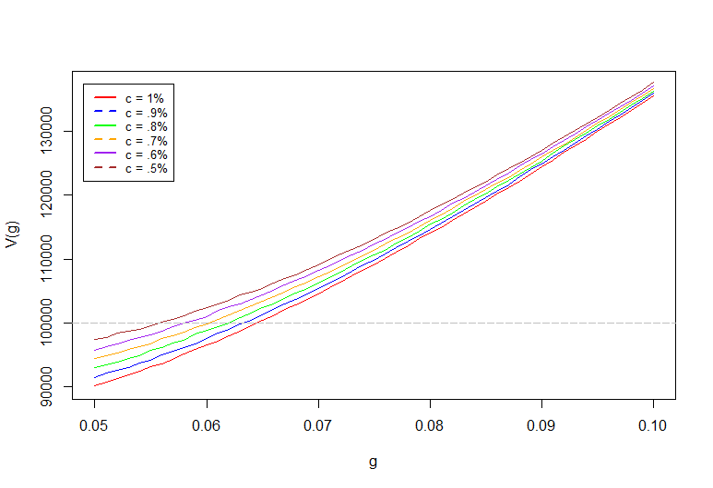}
  \caption{$\sigma$ = 10\%}
\end{subfigure}
\begin{subfigure}{\textwidth}
  \centering
  \includegraphics[width=.5\linewidth]{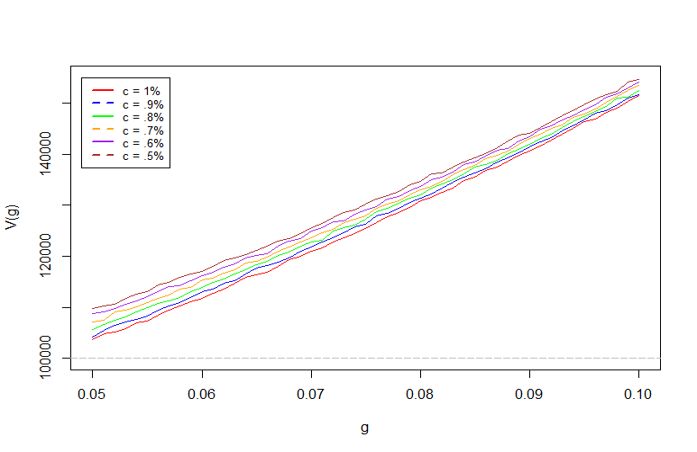}
  \caption{$\sigma$ = 20\%}
\end{subfigure}
\caption{V(g) values for different volatility levels.}
\label{fig:value_givenvol}
\end{figure}

Similarly, Figure \ref{fig:value_givenvol} shows the value of $V(g)$ at each volatility level. The fair value of $g$ varies significantly based on different volatility levels. At a low volatility level of 2\%, the fair value of $g$ is nearly the same for every value of $c$ since the $V(g)$ functions appear to converge to a line. At a level of 10\%, there is more variation amongst $c$ values which results in different and lower fair values of $g$. Lastly, at a volatility level of 20\%, the fair rate $g$ does not even lie between .05 and .1. This reflects the uncertainty in the market, resulting in much lower fair rates which are not competitive. This also suggests that fair rates (and fees) should be linked to the market volatility. The variation caused by the change in volatility shows that the volatility assumption has a very large effect on the results given in sections \ref{subsec:prob} and \ref{subsec:fair_g}.

\subsection{\textit{Rationality of exercising a reset option}}

\subsubsection{Exercise criteria}
\begin{figure}[H]
  \centering
  \includegraphics[width=.75\linewidth]{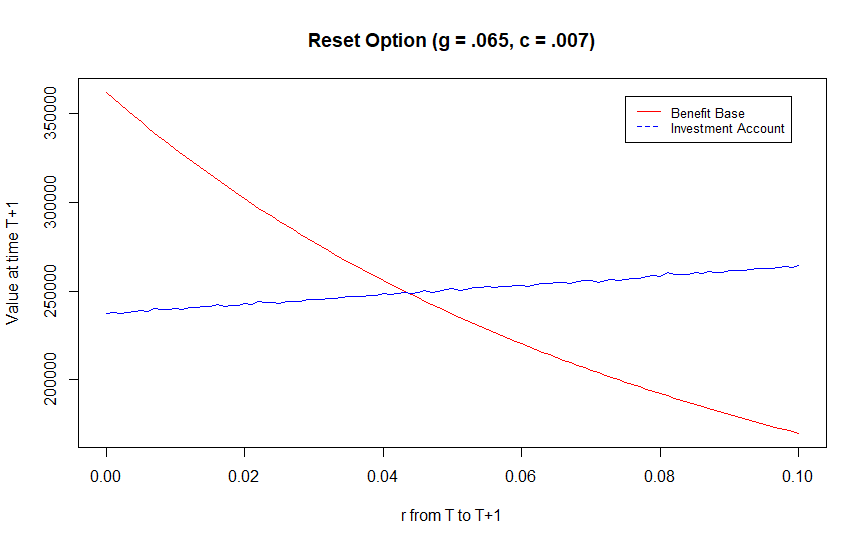}
\caption{Values of $BB(T')$ and $S_f(T')$}
\label{fig_rationality}
\end{figure}

\tab In order to assess the rationality of exercising a reset option as defined by the ability to delay the annuitization date, in this case, by one year, simulations are made for both the performance of the benefit base and the investment account for the year following the initial annuitization date. The performance of these accounts is considered under a new interest rate value (which is the growth rate for the investment account). This value is still constant but varies from 0\% to 10\%. Figure \ref{fig_rationality} shows the projected values for the benefit base and the investment account at time $T'=T+1$ at different interest rate levels. The benefit base rolls-up to a fixed level, so iterations are not run for its value.

\pagebreak
It can be seen that there is a certain interest rate value for the year following the initial annuitization date such that the value of the benefit base equals that of the investment account. This value is defined earlier as the critical interest rate value $r^*$. It is with this value that the rationality of exercising the reset option (at time $T$) is determined. For example, in Figure \ref{fig_rationality} the values of the future benefit base and investment account at time $T'$ are given for an assumed level of $g = .065$ and $c = .007$. If the policyholder was in a situation where they had more money in their investment account than their benefit base at time $T$, then they should rationally exercise the reset option if they expect future interest rates to drop below $r^* = 4.35\%$. This is because if interest rates drop below this level for the following year, the benefit base is now expected to have more value than the investment account. Likewise, if at time $T$ the benefit base has more value than the investment account, the reset option should be rationally exercised if the policyholder expects interest rates to be above 4.35\%.

This can be more generally stated as if $S_f(T) > BB(T)$, then the reset option should be exercised if the policyholder expects future interest rates to be below $r^*$. Likewise, if $S_f(T) < BB(T)$, then the reset option should be exercised if the policyholder expects future interest rates to be above $r^*$. The critical interest rate value stated above of 4.35\% was specifically for $c = .007$ and $g = .065$. What value this critical interest rate has for different levels of $c$ and $g$ is given in the next section.

\pagebreak
\subsubsection{Critical Interest Rate Values}

\begin{figure}[H]
  \centering
  \includegraphics[width=0.75\linewidth]{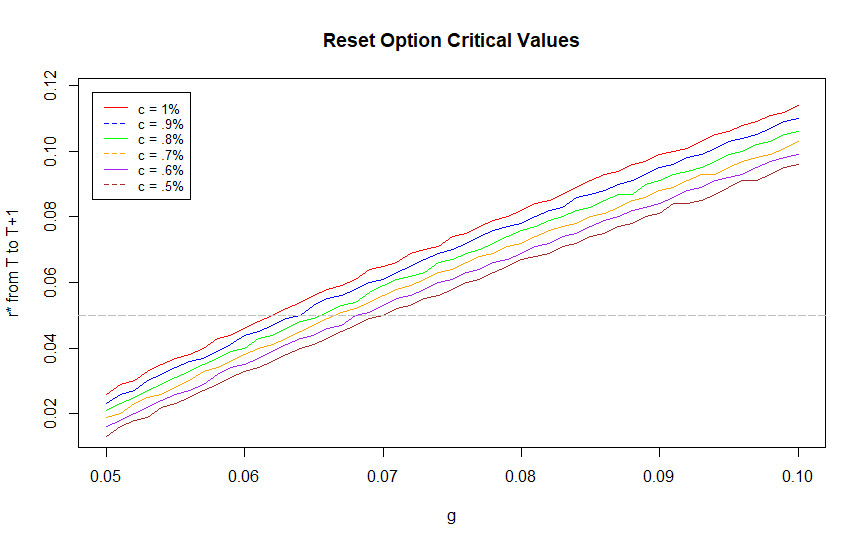}
\caption{Critical values for the interest rate from $T$ to $T'$ ($r^*$) }
\label{fig:critical_r}
\end{figure}

To plot the critical interest rate value for different levels of $c$ and $g$, the data for the chart shown in the previous section is replicated for 306 different combinations of $c$ and $g$. The critical interest rate value $r^*$ is taken from each of these calculations and placed onto a graph as seen in Figure \ref{fig:critical_r}.

The number of iterations that went into calculating the future investment account value at each interest rate level within the individual 306 combinations is again 200,000. This method was computationally heavy and required a lot of computing power and time.
A regression function can be run so that the critical interest rate value can be calculated for any value of $g$. While a linear function could be fit to the data, a second degree polynomial function is a better fit for each level of $c$ between $0.5\%$ and $1\%$. A statistically significant p-value from an ANOVA test verifies this. Additionally, a very high adjusted $R^2$ value tells us that it is not necessary to explore fits for functions with higher degree polynomials. The regression functions are given in Table \ref{table:critical_r}.

\begin{table}[ht]
\begin{center}
\caption{Estimated Value of Critical Interest Rate}
\label{table:critical_r} 
\begin{tabular}{|c|c|c|}
\hline
\multicolumn{1}{|c|}{$c$} &  \multicolumn{1}{c|}{Regression Function} &
\multicolumn{1}{c|}{Adjusted $R^2$}\\
\hline
0.5\% &   $-5.50g^2 + 2.46g - .0953$ & 0.9996\\
0.6\% &   $-5.58g^2 + 2.49g - .0943$ & 0.9997\\
0.7\% &   $-5.96g^2 + 2.56g - .0944$ & 0.9997\\
0.8\% &   $-5.64g^2 + 2.54g - .0916$ & 0.9997\\
0.9\% &   $-5.51g^2 + 2.55g - .0898$ & 0.9997\\
1\%   &   $-5.65g^2 + 2.59g - .0889$ & 0.9997\\
\hline
\end{tabular}
\end{center}
\end{table}

\section{Conclusion} \label{sec:conclusion}

\tab In this paper, we studied the value of a GMIB as a function of the guaranteed annual payment rate $g$ for different level of the fee rate $c$ and the rationality of exercising a reset option based on future expectations of the interest rate. Results have been given to determine the probability of exercising the benefit base, the fair value of $g$, as well as how these quantities are affected by the fee structure and the volatility. Our results show that $g$ has a larger impact on the probability of exercising the benefit base than $c$ does. Assuming financial market parameters set to $\sigma=10\%$ and $r=5\%$, critical $g$ values for fee levels of .5\%, .6\%, .7\%, .8\%, .9\% and 1\% were given to be .0558, .0581, .0601, .0619, .0633, and .0645, respectively. Discussion was provided into how the fee structure and the volatility affect the results of the model. As compared to the alternative fee structure $f_2(n)$, fair levels of $g$ are larger under $f_1(n)$. This is due to an upward shift of the guarantee value from $f_1(n)$ to $f_2(n)$. We also found that the volatility significantly impacts the fair rates. As expected, the lower the volatility the larger the fair rate. In fact, we found that for $\sigma=0.2$, the fair rate falls below the range typically offered in the industry, which suggests that the model parameters $c$ and $g$ are strongly linked to the market volatility.
Lastly, regression functions were produced to determine the critical interest rate value for any value of $g$. These critical interest rate values were used to determine the rationality of exercising the reset option. Future work involves relaxing the assumption of constant interest rates, and incorporating the reset option into the pricing framework.






\end{document}